\documentclass[twocolumn,showpacs,preprintnumbers,amsmath,amssymb]{revtex4}

\usepackage{color}
\usepackage{graphicx}
\usepackage{dcolumn}
\usepackage{bm}

\begin{document}

\preprint{APS/123-QED}

\title{Heavy Fermion Superconductivity in the Quadrupole Ordered State of PrV$_2$Al$_{20}$}

\author{Masaki Tsujimoto}
\author{Yosuke Matsumoto}
\author{Takahiro Tomita}%
\author{Akito Sakai}
\author{Satoru Nakatsuji}
\email{satoru@issp.u-tokyo.ac.jp}

\affiliation{%
Institute for Solid State Physics, University of Tokyo, Kashiwa, Chiba 277-8581, Japan\\
}%

\date{\today}

\begin{abstract}
PrV$_2$Al$_{20}$ is a rare example of a heavy fermion system based on strong hybridization between conduction electrons and nonmagnetic quadrupolar moments of the cubic $\Gamma_3$ ground doublet. Here, we report that a high-quality single crystal of PrV$_2$Al$_{20}$ exhibits superconductivity at $T_{\rm c}=$ 50 mK in the antiferroquadrupole-ordered state under ambient pressure. The heavy fermion character of the superconductivity is evident from the specific heat jump of $\Delta C/T \sim 0.3$ J/mol K$^2$ and the effective mass $m^*/m_0 \sim 140$ estimated from the temperature dependence of the upper critical field. Furthermore, the high-quality single crystals exhibit double transitions at $T_{\rm Q}$ = 0.75 K and $T^{*}$ = 0.65 K associated with quadrupole and octapole degrees of freedom of the $\Gamma_3$ doublet. In the ordered state, the specific heat $C/T$ shows a $T^3$ dependence, indicating the gapless mode associated with the quadrupole and/or octapole order. The strong sensitivity to impurity of the superconductivity suggests unconventional character due to significant quadrupolar fluctuations.
\end{abstract}

\maketitle
$4f$ electron systems exhibit a large variety of nontrivial ground states by tuning the hybridization between localized $4f$ and conduction ($c-$) electrons. Especially, emergence of exotic superconductivity (SC) with a large effective mass from unconventional quantum criticality has attracted much attention \cite{steglich79,Mathur98,CeRhIn5,CeCoIn5,nakatsuji08}. While most of the examples have been reported at the border of magnetism \cite{lohneysen2007,Gegenwart}, similarly exotic states of matter may be found in the vicinity of quantum phase transition associated with different degrees of freedom of $f$-electrons, such as electrical quadrupole (orbital) and valence. In particular, experimental exploration of the quadrupolar instability is important since few studies on the associated quantum criticality has been made to date.

For the exploration, the simplest example could be found in materials that carry no magnetic but quadrupole moments. Such a nonmagnetic ground state is known as the $\Gamma_3$ doublet in the cubic crystalline electric field (CEF) of a $f^2$ configuration. Intensive studies have revealed various interesting states in the cubic $\Gamma_3$ doublet systems, in particular in Pr-based intermetallic compounds \cite{Morin1982257,RefWorks:364,Onimaru,Onimaru2,Onimaru3}. However, Pr-based cubic $\Gamma_3$ doublet systems usually have well-localized quadrupole moments, and thus until quite recently there have been no study on the quadrupolar instability by tuning the $c$-$f$ hybridization.

Pr$Tr_2$Al$_{20}$ ($Tr$ = Ti, V) has been reported as a rare example of cubic $\Gamma_3$-doublet based Kondo lattice systems where one may tune the hybridization strength between quadrupole moments and conduction electrons by chemical substitution and by pressure \cite{akito,Matsubayashi}. These materials have the 
nonmagnetic $\Gamma_3$ CEF ground state with the well separated excited state at $\Delta_{\rm CEF} \sim$ 60 K (Ti) and 40 K (V), as confirmed by various experiments \cite{akito,sato,Nakanishi}, and exhibit the respective ferro- and antiferro- quadrupole ordering at $T_{\rm Q}$ = 2.0 K (Ti) and 0.6 K (V) \cite{akito,sato,Ito}.
Significantly, strong $c$-$f$ hybridization is evident from a number of 
phenomena, including the Kondo effect in the resistivity, Kondo resonance peak observed near the Fermi energy, and the large hyperfine constant in NMR measurements \cite{akito,Matsunami,tokunaga}.
In addition, the hybridization can be enhanced by chemical substitution. In fact,  PrV$_2$Al$_{20}$ exhibits chemical pressure effects such as the enhanced Kondo effect, a large Weiss temperature, and the suppression of the quadrupolar ordering, compared to the Ti analog. Besides, in contrast with the highly localized properties of PrTi$_2$Al$_{20}$, PrV$_2$Al$_{20}$ exhibits non-Fermi liquid behavior above $T_{\rm Q}$ \cite{akito}, which may be attributed to the quadrupolar Kondo effects \cite{Cox}.

Recent discovery of the heavy fermion superconductivity in the vicinity of the putative quadrupolar quantum critical point (QCP) has clearly demonstrated the tunable strong hybridization in PrTi$_2$Al$_{20}$ \cite{Matsubayashi,AkitoSC}. At ambient pressure, PrTi$_2$Al$_{20}$ superconducts at $T_{\rm c}=0.2$ K in the ferroquadrupole ordered state \cite{AkitoSC}. Effective mass estimated from 
the upper critical field is moderately enhanced with $m^*/m_0\sim 16$. Application of pressure significantly increases $T_{\rm c}$ and $m^*$ up to $T_{\rm c} \sim$ 1 K and $m^*/m_0\sim 110$ at $\sim 8$ GPa, while suppressing $T_{\rm Q}$, indicating the emergence of the heavy fermion SC in the vicinity of the QCP of the quadrupolar order\cite{Matsubayashi}. 

In this Letter, we report the first observation of heavy fermion SC in a cubic $\Gamma_3$ doublet compound under ambient pressure. In particular, we discover the superconductivity in an antiferroquadrupolar state of PrV$_2$Al$_{20}$ below $T_{\rm c} =$ 0.05 K in a high-quality single crystal. Both the field dependence of the upper critical field and the specific heat measurements indicate the enhanced effective mass as high as $m^*/m_0\sim 140$ and the specific heat coefficient $\gamma \sim 0.9$ J/mol K$^2$.  Our study using high quality single crystals also revealed the double transition associated with the $\Gamma_3$ doublet, suggesting that not only quadrupolar but octapolar degrees play important roles. 

\begin{figure}[t]
\begin{center}
\includegraphics[keepaspectratio, scale=0.6]{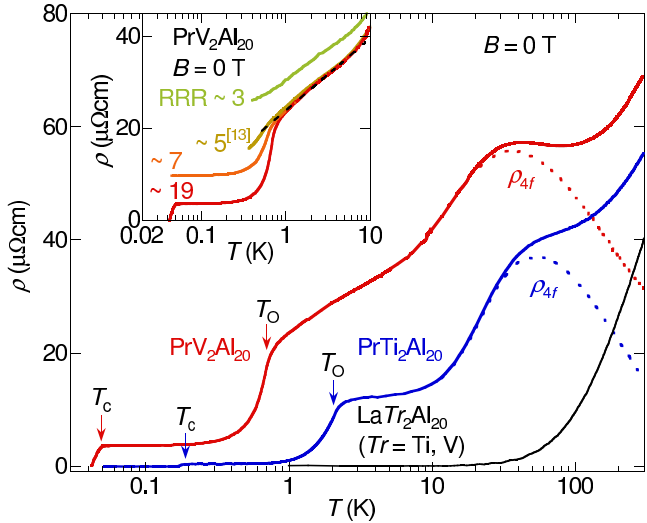}
\caption{(color online) Temperature dependence of the resistivity $\rho(T)$ for the single crystals of PrV$_2$Al$_{20}$ (RRR = 19) and PrTi$_2$Al$_{20}$  (RRR = 150)  under the earth field. $4f$ electron contribution $\rho_{4f}$ is calculated by subtracting the $T$ dependence of the inelastic part of $\rho$ for La$Tr_2$Al$_{20}$ indicated by the solid line. Arrows indicate the peak temperatures found in $d\rho /dT$, $T_{\rm O}=$ 0.68 K (V) and 2.0 K (Ti), and the superconducting transition at $T_{\rm c}=$ 0.05 K (V) and 0.2 K (Ti) , respectively. Inset: $\rho(T)$ of PrV$_2$Al$_{20}$ below 10 K for various crystals with different quality, including the crystal with RRR $\sim$ 19 in the main panel. A fit to a $\ln T$ curve is shown as a broken line at $T > T_{\rm O}$.
}\label{fig1}
\end{center}
\end{figure}

The success in growing high quality single crystals has allowed us to observe the superconductivity in PrV$_2$Al$_{20}$.The details of the experimental method are in the supplemental materials\cite{supplemental}. Figure 1 shows the temperature dependence of the resistivity $\rho (T)$ for one of the highest quality crystals 
with RRR$\sim19$. For comparison, we also plot $\rho(T)$ for PrTi$_2$Al$_{20}$. Clearly, PrV$_2$Al$_{20}$ exhibits a resistivity drop due to the SC transition at $T_{\rm c}$ = 0.05 K as indicated by an arrow. The other arrows indicate the peak temperatures $T_{\rm O}$ found in $d\rho /dT$ associated with quadrupolar ordering. A single peak in $d\rho /dT$ at $T_{\rm O}= 0.68$ K higher than the previous report \cite{akito} was observed for PrV$_2$Al$_{20}$, while clear double transitions at $T_{\rm Q}$ = 0.75 K and $T^{*}$ = 0.65 K were found in the specific heat measurements using the same single crystal as we will discuss later.
The dotted lines indicate $4f$-electron contribution to the resistivity $\rho_{4f}$ calculated by subtracting off the inelastic part of the resistivity of La$Tr_2$Al$_{20}$ ($Tr$ = Ti, V) shown by the solid line in Fig. 1. $\rho_{4f}\propto -\ln T$ observed above $T_{\rm peak}$ $\sim 60$ K (Ti) and $\sim 40$ K (V) should be the magnetic Kondo effect as the magnetic excited CEF states are populated at $T>T_{\rm peak}\sim \Delta_{\rm CEF}$\cite{akito}. 

\begin{figure}[t]
\begin{center}
\includegraphics[keepaspectratio, scale=0.6]{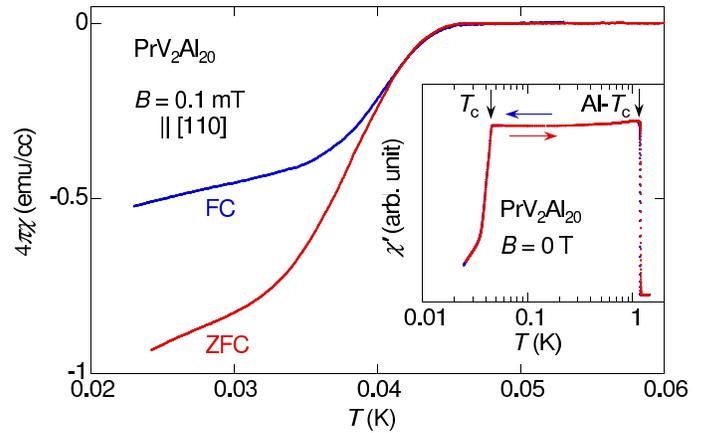}
\caption{(color online) Temperature dependence of the dc-susceptibility $\chi(T)$ for the single crystal of PrV$_2$Al$_{20}$ (RRR $\sim 20$) under the field of 0.1 mT for zero field cooled (ZFC) and field cooled (FC) sequences. Inset: Real part of the ac-susceptibility $\chi'(T)$ for PrV$_2$Al$_{20}$ and the reference of Al.
}\label{fig2}
\end{center}
\end{figure}

To verify the bulk superconductivity, we measured dc- and ac-magnetic susceptibility of a high quality single crystal of PrV$_2$Al$_{20}$(RRR $\sim 20$). Figure 2(a) shows the $T$ dependence of the dc-susceptibility $\chi(T)$ of PrV$_2$Al$_{20}$ under a field of $\mu_0 H = 0.1$ mT along the [110] direction. The unit of $\chi(T)$ is calibrated by using the perfect diamagnetic signal of the reference Al sample with the same dimension placed in the canceling coil. By taking care of the diamagnetic correction factor $D\sim 0.17$ calculated using the rectangular sample dimension\cite{demag}, the volume fraction was estimated to be 82$\%$ (ZFC) and 47$\%$ (FC), indicating the bulk superconductivity. The large diamagnetic signal comparable to the Al shielding signal is also observed in the $T$ dependence of the real part of the ac susceptibility $\chi'(T)$ under zero dc-field and ac-field $\sim0.1$ $\mu$T, as shown in the inset of Fig. 2. 
Bulk SC was also confirmed by the specific heat measurements as we will discuss. 

\begin{figure}[t]
\begin{center}
\includegraphics[keepaspectratio, scale=0.6]{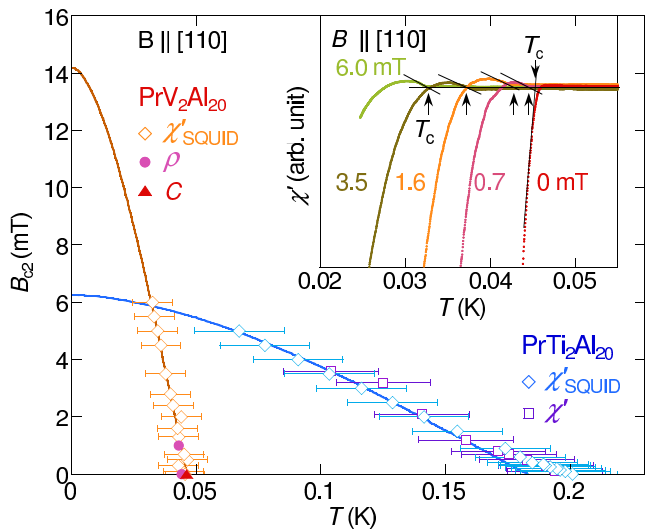}
\caption{(color online) $T$ dependence of the upper critical field $B_{\rm c2}$ for the single crystals of PrV$_2$Al$_{20}$ (RRR $\sim 20$) and PrTi$_2$Al$_{20}$ (RRR = 150) under a filed along [110]. The square/diamond/circle/triangle data points are determined by ac-susceptibility, ac-susceptibility by SQUID, resistivity, and specific heat measurements, respectively. Solid lines represent the fit based on the WHH model. Inset: $T$ dependence of the real part of the ac-susceptibility $\chi'$ for PrV$_2$Al$_{20}$ under various magnetic field. Arrows indicate the critical temperatures.
}\label{fig3}
\end{center}
\end{figure}

To clarify the boundary of the SC phase, we measured the susceptibility, and resistivity under various fields. As a summary, Figure 3 shows the $T$ dependence of the upper critical field $B_{\rm c2}$ for PrV$_2$Al$_{20}$. For comparison, we show the results for PrTi$_2$Al$_{20}$ as well \cite{AkitoSC}. Here, the critical temperatures (diamond) are determined by the onset of the anomaly of the susceptibility defined as the foot of the peak as shown by arrows in Fig. 3 inset. Compared to PrTi$_2$Al$_{20}$, the peak in the susceptibility around $T_{\rm c}$ due to differential paramagnetic effect is quite small in PrV$_2$Al$_{20}$\cite{DPE}, indicating strong pining effect typical to the type II SC. The critical temperatures defined above are consistent with the zero-resistance temperature (circle) of the SC drop of $\rho(T)$ (Fig. 3). The solid line in Fig. 3 is the fit to our $B_{\rm c2}$ results based on the Werthamer-Helfand-Hohenberg (WHH) model \cite{WHH,HW}. The best fitting was obtained using parameters of $T_{\rm c}=46.2$ mK and the slope of $B_{\rm c2}$ at $T_{\rm c}$, $B'_{\rm c2}\equiv dB_{\rm c2}/dT=$ 0.41 T/K. The model reproduces the experimental data well, indicating the orbital depairing effect is dominant. 
The resultant orbital critical field at $T=0$, $B_{\rm c2}^{\rm orb}(0) = -0.727 B_{\rm c2}^{\prime}T_{\rm c}$, and Ginzburg-Landau (GL) coherence length, $\xi=\sqrt{\Phi_0/2\pi B_{\rm c2}^{\rm orb}(0)}$,  are $B_{\rm c2}^{\rm orb}(0) = 14.3$ mT and $\xi = 0.15$ $\mu$m, respectively.

Strikingly, $B'_{\rm c2}$, of PrV$_2$Al$_{20}$ is about 10 times larger than the Ti analog \cite{AkitoSC}, indicating significantly heavier effective mass. 
Indeed, the effective mass is estimated to be $m^*=\hbar k_{\rm F}/ v_{\rm F}\sim140 m_0$ by using
the GL coherence length $\xi = 0.15$ $\mu$m, the Fermi velocity $v_{\rm F}=\xi k_{\rm B}T_{\rm c}/(0.18\hbar)=5.1$ km/s, $k_{\rm F}=(3\pi^2Z/\Omega)=6.1\times 10^9$ 1/m, where $Z$ is the number of electrons per unit cell, and $\Omega$ is the unit-cell volume. 
The effective mass $m^*/m_0\sim 140$ is much larger than $m^*/m_0 \sim 16$ estimated for PrTi$_2$Al$_{20}$ under ambient pressure \cite{AkitoSC},  is comparable to $m^*/m_0\sim 110$ under $\sim 8$ GPa in the vicinity of the quadrupolar quantum criticality \cite{Matsubayashi}. 
Thus, the mass enhancement in PrV$_2$Al$_{20}$ indicates not only the strong $c$-$f$ hybridization, but its proximity to a quadupolar QCP.

\begin{figure}[t]
\begin{center}
\includegraphics[keepaspectratio, scale=0.6]{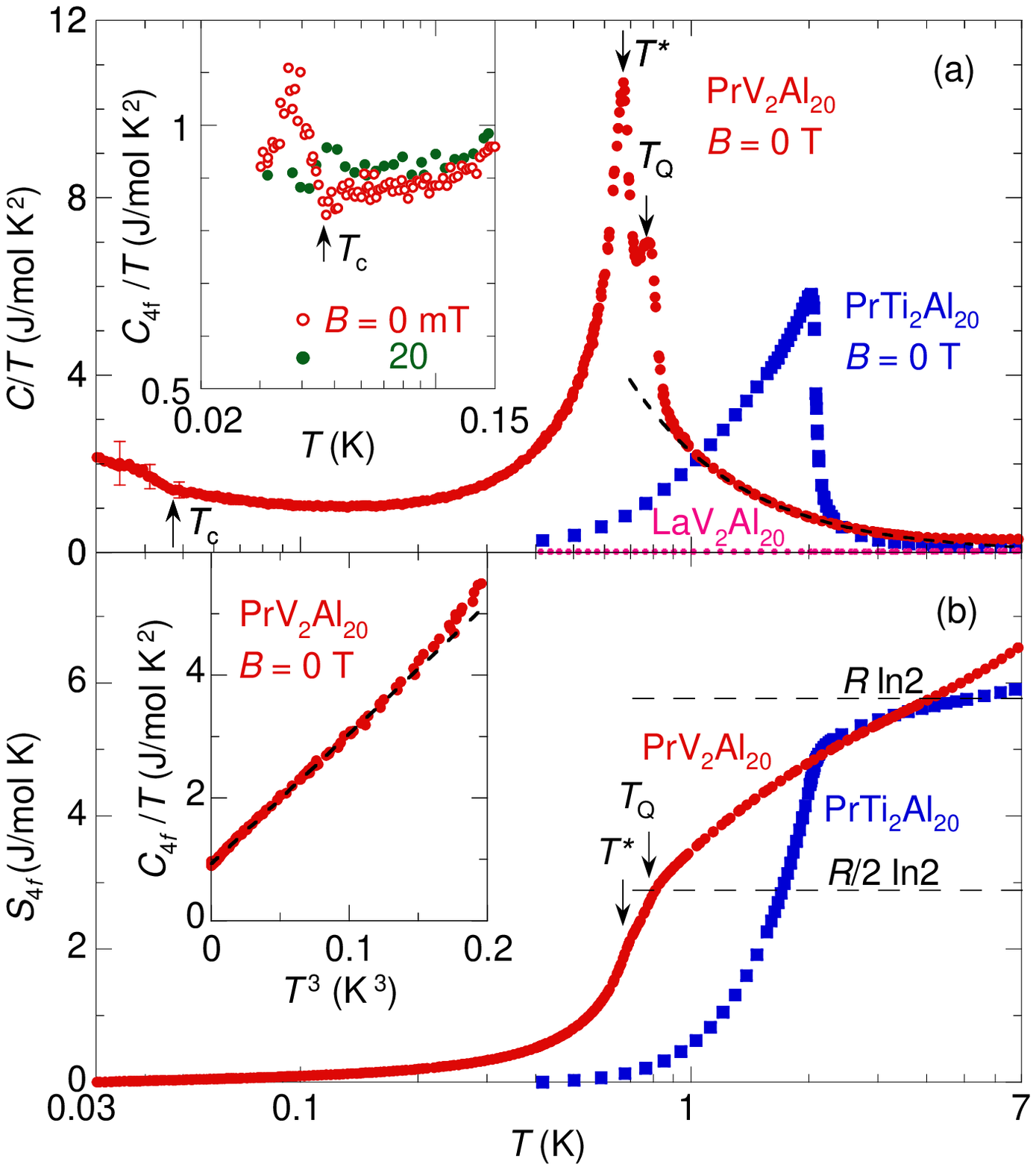}
\caption{(color online) (a) $T$ dependence of the specific heat divided by $T$, $C/T$  for PrV$_2$Al$_{20}$ and PrTi$_2$Al$_{20}$ under the earth field. Double transition temperatures $T_{\rm Q}$ and $T^*$ are defined at the peaks. Broken line indicates the fit to $C/T \sim 1/T^{-3/2}$ in the paraquadrupolar state. As for the error-bars at $T < T_{\rm c}$, see \cite{supplemental}. Inset:  $T$ dependence of $C_{4f}/T$ for PrV$_2$Al$_{20}$ under $B$ = 0 (open circle), 20 mT (closed circle). $C_{4f}/T$ was derived by subtracting the contribution of the lattice and nuclear magnetism from $C/T$. (b)  $T$ dependence of the entropy $S_{4f}$ for PrV$_2$Al$_{20}$ and PrTi$_2$Al$_{20}$. The horizontal broken lines show the value of $R \ln 2$ and $R/2 \ln 2$, respectively. The inset shows $C_{4f}/T$ vs. $T^3$ for the $T$ range at $T <$ 0.58 K. The broken line represents the linear fit, indicating that $C_{4f}/T$ shows $T^3$ dependence in 0.05 K $<$ $T$ $<$ 0.5 K.
}\label{fig4}
\end{center}
\end{figure}

The heavy fermion character of the superconductivity was also confirmed by the specific heat measurements.  
Figure \ref{fig4} (a) shows the specific heat divided by temperature $C/T$ of  PrV$_2$Al$_{20}$.  In comparison, the lattice contribution estimated from $C/T$ of LaV$_2$Al$_{20}$ is found small and negligible. 
After showing a broad minimum at $T\sim 0.12$ K, $C/T$ slightly increases on cooling and exhibits an anomaly at $T=$ 0.046 K, corresponding to the SC transition. 

The low $T$ upturn in $C/T$ becomes evident in the normal state stabilized under the magnetic field of 20 mT, and is found to follow $C/T \sim 1/T^{2}$ down to the lowest $T$ of 30 mK.
This power law increase, $C \sim 1/T$, seen at $ T < 100$ mK indicates an entropy release associated with a very small energy scale of mK range, most likely coming from a nuclear magnetism. Actually, hyperfine-enhanced nuclear magnetism is often reported for Pr intermetallic compounds with a nonmagnetic ground state. Indeed, the analysis based on the $\mu$SR measurements indicated the hyperfine-enhanced nuclear magnetism also in PrTi$_2$Al$_{20}$ and estimated the transition temperature of 0.13 mK \cite{Ito}. In addition, $C \sim 1/T^\alpha (\alpha \sim 1)$ behavior has been commonly seen  in the $T$ range close to the nuclear magnetic transition. For example, PrNi$_5$, a prototypical system of the hyperfine-enhanced nuclear magnetism exhibits $C \sim 1/T^\alpha (\alpha \sim 1)$ behavior in the similar $T$ range below 100 mK down to near the ordering temperature of 0.4 mK \cite{PrNi5}.


Thus, we estimated the 4$f$ electron contribution of the specific heat divided by $T$, $C_{4f}/T$ by subtracting the low $T$ upturn ($\sim 1/T^2$) coming from the nuclear magnetism, as shown in the inset of Fig. \ref{fig4} (a). The zero field SC anomaly is evident indicating its bulk character, which can be suppressed under the field of 20 mT. The nearly constant $C_{4f}/T$ in the normal state provides an estimate of  the large electronic coefficient $\gamma\sim 0.9$ J/molK$^2$,
consistent with the above estimate of the effective mass $m^*/m_0\sim 140$.
From the sudden increase of $C_{4f}/T$ at $T_{\rm c}$, one may estimate the SC jump in $\gamma$, $\Delta C/T_{\rm c}\sim 0.3$ J/mol K$^2$, which provides direct evidence of the heavy fermion SC. This yields the ratio $\Delta C/(\gamma T_{\rm c}) \sim 0.3$, which is much smaller than the BCS value of 1.43. 
It is an interesting future issue to check if the transition becomes much sharper as the sample quality is further improved.


In the normal state, we observed clear two peaks of the specific heat at $T_{\rm Q}=$ 0.75 K and $T^*=$ 0.65 K, indicating double transitions, as discussed by K. Araki {\it et al}  in Ref. \cite{Araki} (Fig. 4(a)). The double transition was not found in the sample with RRR = 6 \cite{akito}, but observed only in relatively high RRR ($> 7$) samples\cite{Araki}, and thereby an intrinsic phenomenon. Given the fact that  both quadrupole and octapole degrees of freedom are available in the $\Gamma_3$ doublet state and may not necessarily coexist with each other, one interesting possibility is that the high and low temperature transitions are due to the octapole and quadrupole ordering, respectively.

Strong hybridization between quadrupole moments and conduction electrons not only is the key to understand the heavy fermion character of the superconductivity, but also induces various interesting effects in the normal state. 
A hybridization effect is seen in the temperature dependence of the entropy $S(T)$ obtained after integration of $C_{4f}/T$ vs. $T$.
As shown in Fig. 4(b), $S(T)$ of  PrV$_2$Al$_{20}$ reaches $R\ln2$ around 4 K $\sim 6 T_{\rm Q}$ and is further suppressed on cooling down to 50 \% of $R\ln2$ at $T_{\rm Q}$, while the suppression is much weaker and it retains 90 \% at $T_{\rm Q}$ in the Ti analog.
The release of the $\Gamma_3$ entropy over a wide range of $T$ below $6  T_{\rm Q}$  cannot be ascribed to the critical fluctuations associated with the quadrupolar ordering. Instead, it should come from the screening of quadrupole moments by conduction electrons through the strong hybridization. In the same temperature region, we have seen various anomalous metallic properties far different from the Fermi liquid behavior. For example, $C_{4f}/T$ shows unusual increase proportional to $T^{-3/2}$ as shown in Fig. 4 (a), while the magnetic susceptibility exhibits $-T^{1/2}$ dependence \cite{akito}. 
In addition, $\rho(T)$ also largely deviates from the Fermi liquid $\rho(T)\propto T^2$ law and behaves as $\rho(T)\propto \ln T$ (Fig. 1 inset). One possible scenario for the origin of the anomalous metal is the quadrupolar Kondo effect. Interestingly, $S(T)$ becomes exactly $R/2\ln2$ at $T_{\rm Q}$, as expected for the quadrupolar Kondo effect.

As indicated in Fig. \ref{fig1} inset, only the highest quality sample exhibits the sharp drop of the resistivity at the quadrupolar transition as well as the superconductivity. By decreasing RRR, the superconductivity immediately disappears, and the resistivity drop below $T_{\rm Q}$ becomes gradually weak and finally disappears for RRR $< 5$. In particular, the strong sensitivity of SC to the sample quality suggests unconventional character of the SC. On the other hand, the paraquadrupolar state $\rho(T)$ curves at $T > T_{\rm Q}$ overlaps on top for all the samples with RRR $> 5$. This indicates that the inelastic process governing $\rho(T)$ does not depend on the impurity concentration, and thus is local in character. It is most likely that the anomalous metallic state arises from the {\it local screening} of quadrupole moments of the $\Gamma_3$ state through, for example, the quadrupolar Kondo effect.

Strong hybridization also affects the excitation spectrum in the quadrupolar ordered state. Due to the anisotropic character, quadrupole moments normally form a excitation gap in the ordered state. Indeed, PrTi$_2$Al$_{20}$ displays the exponential decay of the specific heat and resistivity below $T_{\rm Q}$ \cite{AkitoSC,SakaiProceeding}. 
In PrV$_2$Al$_{20}$, however, the temperature dependence of $C_{4f}$ below $T_{\rm Q}$ shows the $T^4$ power law behavior. Figure 4(b) inset indicates the $T^3$ dependence of $C_{4f}/T$ in the temperature range between 0.05 K and 0.5 K. This suggests a gapless mode due to the quardrupolar order in PrV$_2$Al$_{20}$, corresponding to ``orbiton'' in transition metal systems \cite{orbiton}.

Although there is no theoretical prediction to the best of our knowledge, the anomalous metallic state in the paraquadrupolar state and the gapless quadrupolar excitations in the ordered state may well come from the competition between the quadrupole order and the screening effects due to strong hybridization, and are therefore the signatures of strong quadrupole fluctuations arising from the proximity to a quadrupolar quantum phase transition. It is highly likely that the pairing of the heavy fermion superconductivity is mediated by such fluctuations. It is thus quite interesting to study the pressure effect of a high quality single crystal of PrV$_2$Al$_{20}$. As PrV$_2$Al$_{20}$ has the stronger hybridization than PrTi$_2$Al$_{20}$, the putative quantum phase transition and the associated heavy fermion superconductivity would emerge at a much lower pressure than in PrTi$_2$Al$_{20}$.

\begin{acknowledgments}
We thank Y. Uwatoko, K. Matsubayashi, J. Suzuki, T. Sakakibara, Y. Shimura, K. Araki, Y-B. Kim, E. C. T. O'Farrell, K. Kuga and K. Ueda for useful discussions. 
This work was partially supported by Grants-in-Aid (No. 25707030) from the Japanese Society for the
Promotion of Science, by Grants-in-Aids for Scientific Research on Innovative Areas ``Heavy Electrons'' of the Ministry of Education, Culture, Sports, Science and Technology, Japan. The use of the facilities of the Materials Design and Characterization Laboratory at the Institute for Solid State Physics, The University of Tokyo, is gratefully acknowledged.
\end{acknowledgments}

\bibliography{bibfile_PrV2Al20_SC}

\end{document}


\preprint{APS/123-QED}

\title{Heavy Fermion Superconductivity in the Quadrupolar Ordered State of PrV$_2$Al$_{20}$: Supplemental Material}

\author{Masaki Tsujimoto}
\author{Yosuke Matsumoto}
\author{Takahiro Tomita}%
\author{Akito Sakai}
\author{Satoru Nakatsuji}
\email{satoru@issp.u-tokyo.ac.jp}

\affiliation{%
Institute for Solid State Physics, University of Tokyo, Kashiwa, Chiba 277-8581, Japan\\
}%

\date{\today}

\maketitle{\bf Experimental method}


Single crystals of $RTr_2$Al$_{20}$ ($R$ = Pr, La, $Tr$ = Ti, V) were grown by the Al self-flux method \cite{akito}. We have succeeded in growing high quality single crystals, whose residual resistivity ratio (RRR) reaches more than 10, by tuning the starting ratio. Single and powder X-ray diffraction and SEM-EDX measurements indicate high quality and single phase of obtained crystals. The electrical resistivity $\rho(T)$ was measured by using a low-frequency ac four-terminal method. In this paper, we define RRR $\equiv \rho(300 {\rm K})/ \rho(0.05 {\rm K})$. 
The ac- and dc-susceptibility measurements were performed by using a home-made magnetometer comprising a commercial SQUID sensor (Tristan Technologies) installed in a dilution refrigerator \cite{SQUID}. 
The weight of the sample used for the SQUID and specific heat measurements is 0.0987 mg and RRR$\sim$ 20. The size is 0.2 mm $\times$ 0.2 mm $\times$ 0.6 mm.
All the ac field ($\sim$ 0.1 $\mu$T, 17 Hz) and dc field ($\le$10 mT) were applied along the [110] direction. As a reference superconductor, a piece of element Al (5N) reshaped to the same dimension as the sample was put in the canceling coil wound in reverse to the pick-up coil. 
The specific heat was measured by a relaxation method for the same sample as used for the ac- and dc-susceptibility measurements. 
Measurements below $T\sim 1$ K were made by using a specific heat cell installed in a $^3$He-$^4$He dilution refrigerator \cite{matsumoto_Cp}. 
For the measurements above  $T=0.9$ K, a commercial system (PPMS, Quantum Design) was used.\\[5mm]

\maketitle{\bf Errors in the heat capacity measurement at very low temperatures}


Rather large error bars of the specific heat measured below 0.1 K in Fig. 4 (a) correspond to a systematic error 
between the data obtained from cooling and warming relaxations.   
This is supposed to originate from a slightly different ambient heat leak to the sample stage during the warming and cooling relaxations ($\Delta\dot{Q}\sim$ 0.6 pW), 
which may affect the both warming and cooling relaxations in an opposite manner. 
The discussion on the SC jump given above is not affected by this error, as the data taken with warming or cooling processes  indicate almost the same SC anomaly.

\bibliography{bibfile_PrV2Al20_SC}